\documentclass[]{jfm}

\usepackage{graphicx}
\usepackage{newtxtext}
\usepackage{newtxmath}
\usepackage{natbib}
\usepackage{hyperref}
\usepackage{xcolor}
\usepackage{amsmath}
\usepackage{amsfonts}

\hypersetup{
    colorlinks = true,
    urlcolor   = blue, 
    citecolor  = black,
}

\newcommand{\RomanNumeralCaps}[1]
\linenumbers

\title{Super-resolution with dynamics in the loss}
\author{Jacob Page\aff{1}\corresp{\email{jacob.page@ed.ac.uk}}}
\affiliation{\aff{1}School of Mathematics, University of Edinburgh, Edinburgh, EH9 3FD, UK}

\begin{document}
\maketitle

\begin{abstract}
Super-resolution of turbulence is a term used to describe the prediction of high-resolution snapshots of a flow from coarse-grained observations.
This is typically accomplished with a deep neural network and training usually requires a dataset of high-resolution images.
An approach is presented here in which robust super resolution can be performed without access to high-resolution reference data, as might be expected in an experiment. 
The training procedure is similar to data assimilation, wherein the model learns to predict an initial condition that leads to accurate coarse-grained predictions at later times, while only being shown coarse-grained observations. 
Implementation of the approach requires the use of a fully differentiable flow solver in the training loop to allow for time-marching of predictions. 
A range of models are trained on data generated from forced, two-dimensional turbulence. 
The networks have reconstruction errors which are similar to those obtained with `standard' super-resolution approaches using high resolution data.
Furthermore, they significantly outperform data-assimilation for state-estimation on individual trajectories, allowing accurate reconstruction on coarser grids than is possible with standard variational approaches. 
\end{abstract}

\section{Introduction}
\label{sec:intro}
The rise of convolutional neural networks (CNNs) in computer vision has spurred a flurry of applications in fluid mechanics \citep{Brunton2020}, where there is natural crossover due to local or global equivariances and the expectation of lower-dimensional dynamics beneath the high-dimensional observations. 
One example of successful crossover from computer vision to fluid mechanics is `super-resolution' \citep{Dong2016} where a CNN is trained to generate a realistic high-resolution image from a low-resolution input. 
This is typically done by showing a neural network many examples of high-resolution snapshots which have been coarse-grained; the network then learns a data-driven interpolation scheme rather than fitting a low-order polynomial \citep{Fukami2019}. 

CNNs trained to perform super-resolution have demonstrated a remarkable ability to generate plausible turbulent fields from relatively scarce observations \citep{Fukami2019,Fukami2020}.
Recent efforts have focused on accurate reproduction of terms in the governing equation, or on the reproduction of spectral properties of the flow \citep{Fukami2023}, which is often attempted with additional terms in the loss function used to train the model.
Almost all examples rely on a dataset of high-resolution ground-truth images, with some recent focus on maintaining accuracy in a low-data limit \citep{Fukami2024}. 
One exception to this is the study from \citet{Kelshaw2022} in low Reynolds-number Kolmogorov flow, where high resolution images were generated by minimising a loss evaluated on the coarse grids, which includes a small contribution from a term encouraging the solution so satisfy the governing equations at each point on the coarse grid.

This paper introduces a robust super-resolution algorithm for high-Reynolds number turbulence which does not require high-resolution data, but instead is trained by attempting to match the time-advanced, super-resolved field to a coarse-grained trajectory.
A key component in this approach is the use of a fully-differentiable flow solver in the training loop, which allows for the time marching of neural network outputs and hence the use of trajectories in the loss function. 
The inclusion of a `solver in the loop' \citep{solverinloop} when training neural networks has been shown to be a particularly effective way to learn new numerical schemes \citep{Kochkov2021} and to learn effective, stable turbulence parameterisations \citep{List2022}. 
With the coarse-grained trajectory comparison used here, the optimisation problem for network training is very similar to variational data assimilation \citep[4DVar -- see][]{Li2019,Wang2021}. 
For 4DVar of homogeneous turbulence, a key bottleneck to accurate reconstruction of the smaller scales is the level of coarse-graining, with a consensus emerging on a critical lengthscale of $l_C \approx 5\pi \eta_K$ \citep{Lalescu2013,Li2019}, where $\eta_K$ is the Kolmogorov scale \citep[note that in wall-bounded flows there are different criteria involving the Taylor microscale][]{Wang2021,Wang2022}. 
The hope with incorporating a neural network in this optimisation process is that, through exposure to a wide range of time evolutions, the model can learn an efficient parameterisation of the inertial manifold and thus generate plausible initial conditions where 4DVar would be expected to struggle. 



%
%
\section{Flow configuration and neural networks}
\label{sec:neural1}

\subsection{Two-dimensional turbulence}
\begin{figure}
    \centering
    \includegraphics[width=\linewidth]{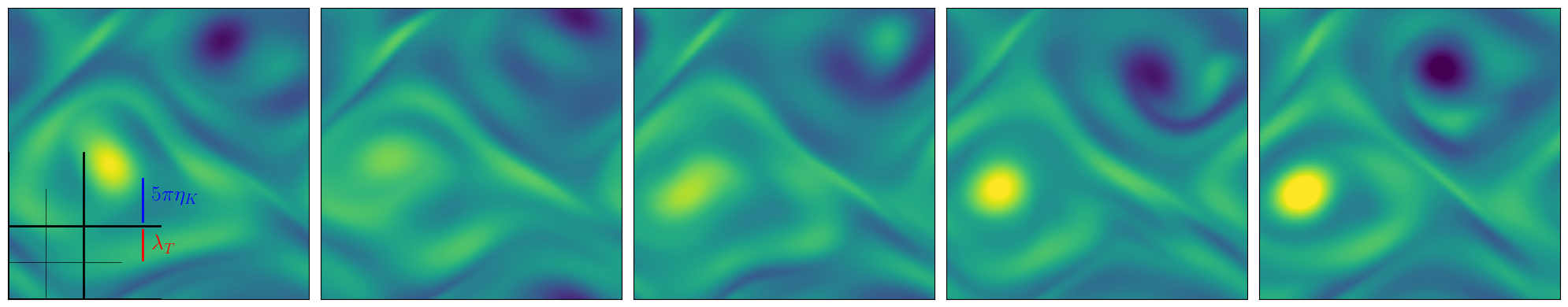}
    \includegraphics[width=\linewidth]{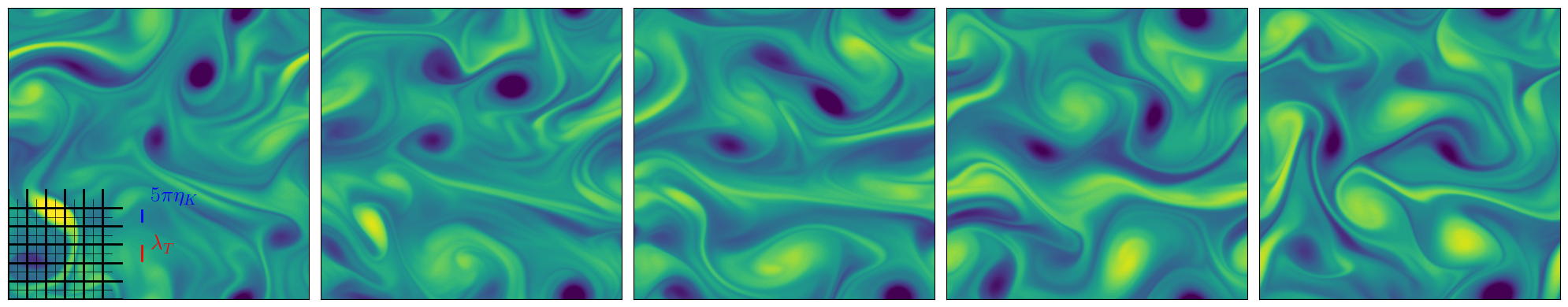}
    \caption{Snapshots of spanwise vorticity from example trajectories at $Re=100$ (top, contours run between $\pm 10$) and $Re=1000$ (bottom, contours run between $\pm 15$). 
    The black grids in the leftmost panels highlight coarse graining by a factor of 16 (thin lines) or 32 (thick lines) relative to the simulation resolution. The Taylor microscale (relative to the lengthscale $L_x^* / (2\pi)$) is indicated by the labelled vertical red lines, 
    while the blue line measures $5\pi \eta_K$.
    }
    \label{fig:vort_examples}
\end{figure}
The super-resolution approach developed in this paper is applied to monochromatically forced, two-dimensional turbulence on the two-torus (Kolmogorov flow). 
Solutions are obtained by time integration of the out-of-plane vorticity equation,
\begin{equation}
    \partial_t \omega + \mathbf u \cdot \boldsymbol{\nabla} \omega = \frac{1}{Re}\Delta \omega - \alpha \omega - n \cos(n y).
\label{eqn:vorticity}
\end{equation}
The velocity $\mathbf u = (u, v)$ may be obtained from the vorticity via solution of the Poisson problem for the streamfunction $\Delta \psi = -\omega$, where $u = \partial_y \psi$ and $v = -\partial_x\psi$. 
Equation (\ref{eqn:vorticity}) has been non-dimensionalised with a lengthscale based on the fundamental wavenumber for the box, $1/k^* = L_x^* / 2\pi$, and a timescale defined as $(k^* \chi^*)^{-1/2}$, where $\chi^*$ is the strength of the forcing per unit mass in the momentum equations.
This results in a definition of the Reynolds number $Re:= (\chi^* / k^{*3})^{1/2} /\nu$.
As in previous work \citep{Chandler2013,Page2021,Page2024} the number of forcing waves is set at $n=4$, while the box has unit aspect ratio such that its size is $L_x = L_y = 2\pi$ in dimensionless units.

Two Reynolds numbers are considered in this work, $Re= 100$ and $Re=1000$. 
For the latter, a linear damping term is included in (\ref{eqn:vorticity}) with coefficient $\alpha=0.1$ to prevent the build up of energy in the largest scales. 
Representative snapshots of the vorticity at both Reynolds numbers are reported in figure \ref{fig:vort_examples}, where the Taylor microscale is also indicated as well as the `critical' lengthscale for assimilation $l_C := 5\pi \eta_K$.
Equation (\ref{eqn:vorticity}) is solved using the fully differentiable \texttt{JAX-CFD} solver \citep{Kochkov2021}. 
The psuedospectral version \citep{Dresdner2022} of this solver is used here, with resolution $N_x \times N_y = 128\times 128$ at $Re=100$ and $N_x \times N_y = 512 \times  512$ at $Re=1000$.

\subsection{Super-resolution problem and neural network architecture}
The objective of super-resolution is to take a coarse-grained flow field, here a velocity snapshot, and interpolate it to a high-resolution grid. 
The coarse-graining operation considered here, denoted $\mathcal C: \mathbb R^{N_x\times N_y\times 2} \to \mathbb R^{N_x/M \times N_y/M\times 2}$, is defined by sampling the original high-resolution data at every $M > 1$ gridpoint in both $x$ and $y$, motivated by observations that may be available in an experiment. 
A function $\mathcal N:  \mathbb R^{N_x/M \times N_y/M\times 2} \to \mathbb R^{N_x\times N_y\times 2}$ is then sought which takes a coarse-grained turbulent snapshot $\mathcal C(\mathbf u)$ and attempts to reconstruct the high-resolution image. 
The function $\mathcal N$ is defined by a convolutional neural network described below, and is found via minimisation of a loss function. 
In `standard' super-resolution \citep{Fukami2019,Fukami2020} this loss usually takes the form
\begin{equation}
    \mathscr L_{SR} = \frac{1}{N_S }\sum_{j=1}^{N_S}\| \mathbf u_j -  \mathcal N_{\boldsymbol \Theta} \circ \mathcal C ( \mathbf u_j) \|^2,
    \label{eqn:standard_sr_loss}
\end{equation}
where $N_S$ is the number of snapshots in the training set and $\boldsymbol{\Theta}$ are the weights defining the neural network.
Note that in this work only reconstruction of velocity is used, due to challenges in estimating vorticity from coarse-grained experimental data. 
Models trained on vorticity outperform the velocity networks in a variety of ways, particular in reproduction of the spectral content (not shown).

Various approaches have added additional terms to equation (\ref{eqn:standard_sr_loss}) to encourage the output to be more physically realistic \citep[e.g. solenoidal, prediction of specific terms in the governing equation -- see][]{Fukami2023}.  
In this work incompressibility is built into the network architecture (see below), while the loss considered here does include additional physics but this is done using trajectories obtained from time marching (\ref{eqn:vorticity}): 
\begin{equation}
    \mathscr L_{TF} = 
    \frac{1}{N_S N_T}\sum_{j=1}^{N_S}\sum_{k=1}^{N_t}\| \boldsymbol \varphi_{t_k}(\mathbf u_j) - \boldsymbol \varphi_{t_k}\circ \mathcal N_{\boldsymbol \Theta} \circ \mathcal C ( \mathbf u_j) \|^2,
    \label{eqn:loss_nn_t}
\end{equation}
where $\boldsymbol{\varphi}_t$ is the time-forward map (with conversion from velocity to vorticity and back) of (\ref{eqn:vorticity}), and 
$t_k \in \{0, \Delta t, \cdots (N_t -1)\Delta t\}$ with $\Delta t = M \delta t$, where $\delta t$ is a simulation timestep.
The unroll time, $T=(N_t - 1)\Delta t$, and coarsening factor $M$ are design choices. 
Here the temporal coarsening is set to match the coarsening in space, $M\equiv 2^n$, with either $n\in \{4, 5\}$.
The selection of the unroll time is constrained by the Lyapunov time for the system. 
In this work $T=1.5$ at $Re=100$, while results will be shown at both $T\in\{0.5, 1\}$ at $Re=1000$.
These choices were guided by the performance of data assimilation at the finer value of $M=16$, shown alongside the super resolution results in \S 3, where reconstruction errors after an unroll time are $O(2\%)$ at $Re=100$ and $O(1\%)$ at $Re=1000$ (with $T=1$).
Training a neural network with a loss like (\ref{eqn:loss_nn_t}) requires backpropagation of derivatives with respect to parameters through the time-forward map. This is accomplished here through use of the fully differentiable flow solver \texttt{JAX-CFD} which was created for this purpose \citep{Kochkov2021,Dresdner2022}. 

While (\ref{eqn:loss_nn_t}) incorporates time evolution, it still requires access to high-resolution reference trajectories. 
An alternative loss, inspired by data assimilation, can be defined based only on coarse observations,
\begin{equation}
    \mathscr L_{TC} = \frac{1}{N_S N_T}\sum_{j=1}^{N_S}\sum_{k=1}^{N_t}\| \underbrace{\mathcal C \circ \boldsymbol\varphi_{t_k}(\mathbf u_j)}_{(i)} - \underbrace{\mathcal C \circ \boldsymbol\varphi_{t_k}\circ \mathcal N_{\boldsymbol \Theta} \circ \mathcal C ( \mathbf u_j)}_{(ii)} \|^2.
    \label{eqn:loss_nn_da}
\end{equation}
The highlighted terms represent: (i) a coarse-grained trajectory -- accomplished here by coarse-graining the output of the flow solver, but conceptually this could also be a set of experimental measurements at a set of sequential times -- and (ii) the forward trajectory of the super-resolved prediction, which is then coarse-grained for comparison to (i).

\begin{figure}
    \centering
    \includegraphics[width=\textwidth]{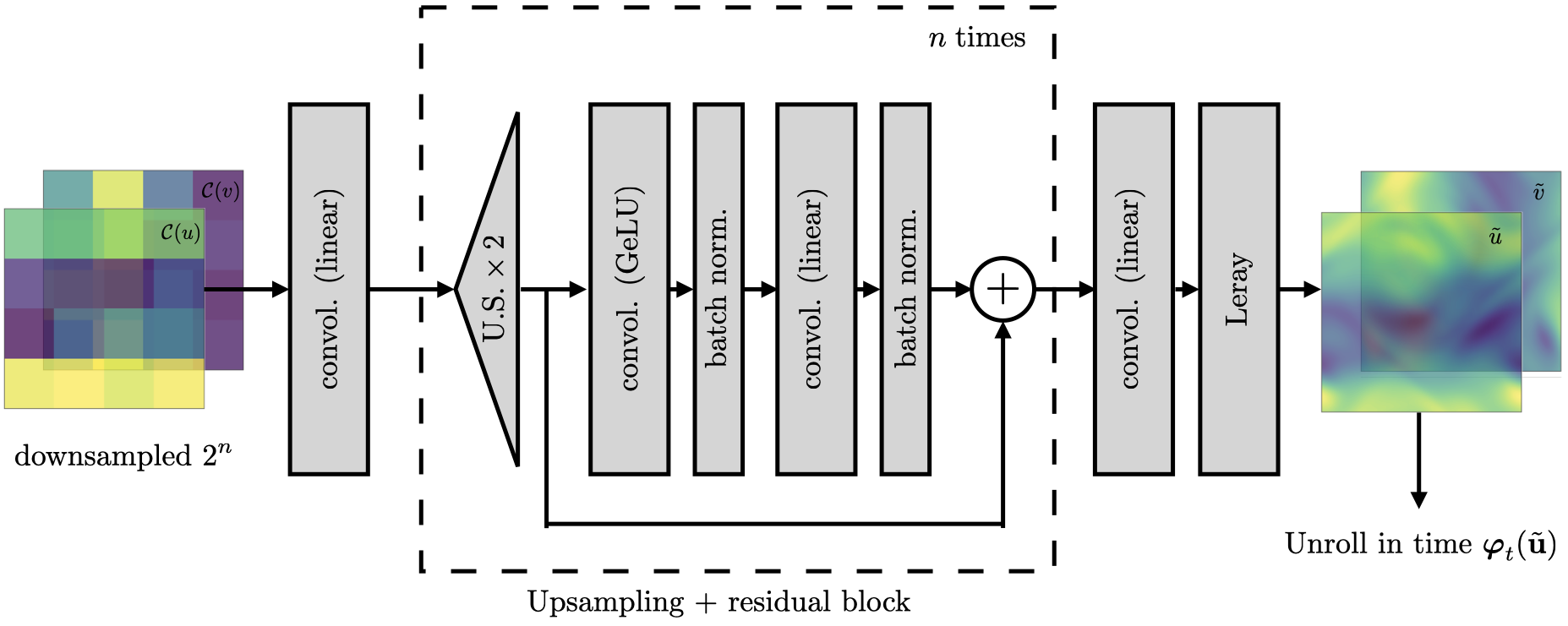}
    \caption{Schematic of the neural network architecture adopted in this study. `U. S.' indicates upsampling. The output of the network is time marched to compute the loss function (equation \ref{eqn:loss_nn_t} or \ref{eqn:loss_nn_da}).}
    \label{fig:schematic}
\end{figure}
The neural networks used in this study are purely convolutional with a `ResNet' structure \citep{He2016}. 
If the coarsening reduces the dimension of the input by a factor of $2^n$, then $n$ residual layers with upsampling by a factor of 2 are used to reconstruct the high-resolution field. 
The output of the $n$ residual convolutions is a field, $\tilde{\mathbf u}'$, with the correct dimensionality (resolution $N_x \times N_y$ and 2 channels for $u$ and $v$).
However, $\tilde{\mathbf u}'$ does not by default satisfy the divergence-free constraint. 
Rather than add a soft constraint to the loss, an additional layer is included in the neural network which performs a Leray projection to produce the divergence free-prediction for the high resolution velocity,
$
    \mathscr P(\tilde{\mathbf u}') = \tilde{\mathbf u}' - \boldsymbol{\nabla} \Delta^{-1} \boldsymbol{\nabla} \cdot \tilde{\mathbf u}'.
$

The architecture is sketched in figure \ref{fig:schematic}.
A fixed number of 32 filters is used at each convolution, apart from the final layers where only 2 channels are required for the velocity components.
The kernel size is fixed at $4\times 4$ across the network and periodic padding is used for the convolutions. 
In this work downsampling is performed by a factor of either $16$ or $32$, which results in a total number of trainable parameters of either $N_{16}\sim 1.33 \times 10^5$ or $N_{32}\sim 1.67 \times 10^5$ respectively. 
While some experimentation was performed with varying the kernel size through the network, the architecture has not been heavily optimised and more complex configurations may be more effective. 
Either GeLU \citep{gelu_arxiv} or linear (no) activation functions are used between layers, as indicated in figure \ref{fig:schematic}. 

The models examined below were trained on relatively small datasets of 60 trajectories at each $Re$, each consisting of 50 snapshots spaced by $\Delta t = 2$ (a transient burn-in portion of each trajectory was discarded prior to saving data). 
Data augmentation of random shift-reflects and rotations was applied during training.
Networks were all trained using an Adam optimiser \citep{Kingma2015} with a learning rate of $\eta = 10^{-4}$ and a batch size of 16 for 50 epochs, where the model with best-performing validation loss (evaluated on 10\% of data partitioned off from the training set) was saved for analysis. 
Results reported in \S 3 were all obtained on separate test datasets generated via the same approach.

Note that the rather large downsampling factors used in the models result in a set of coarse grained observations which are either spaced roughly a Taylor microscale, $\lambda_T$, apart ($M=16$) or spaced further than the microscale ($M=32$).
The same is true for the spacing relative to the `critical' lengthscale, $l_C \approx 5\pi \eta_K$. 
As a result, `standard' data assimilation strategies struggle at both resolutions, which is discussed further below.
The coarsened grids, $\lambda_T$ and $l_C$ were all included earlier in figure \ref{fig:vort_examples} for comparison.

\section{Super-resolution with dynamics}


\begin{figure}
    \centering
    \includegraphics[width=\linewidth]{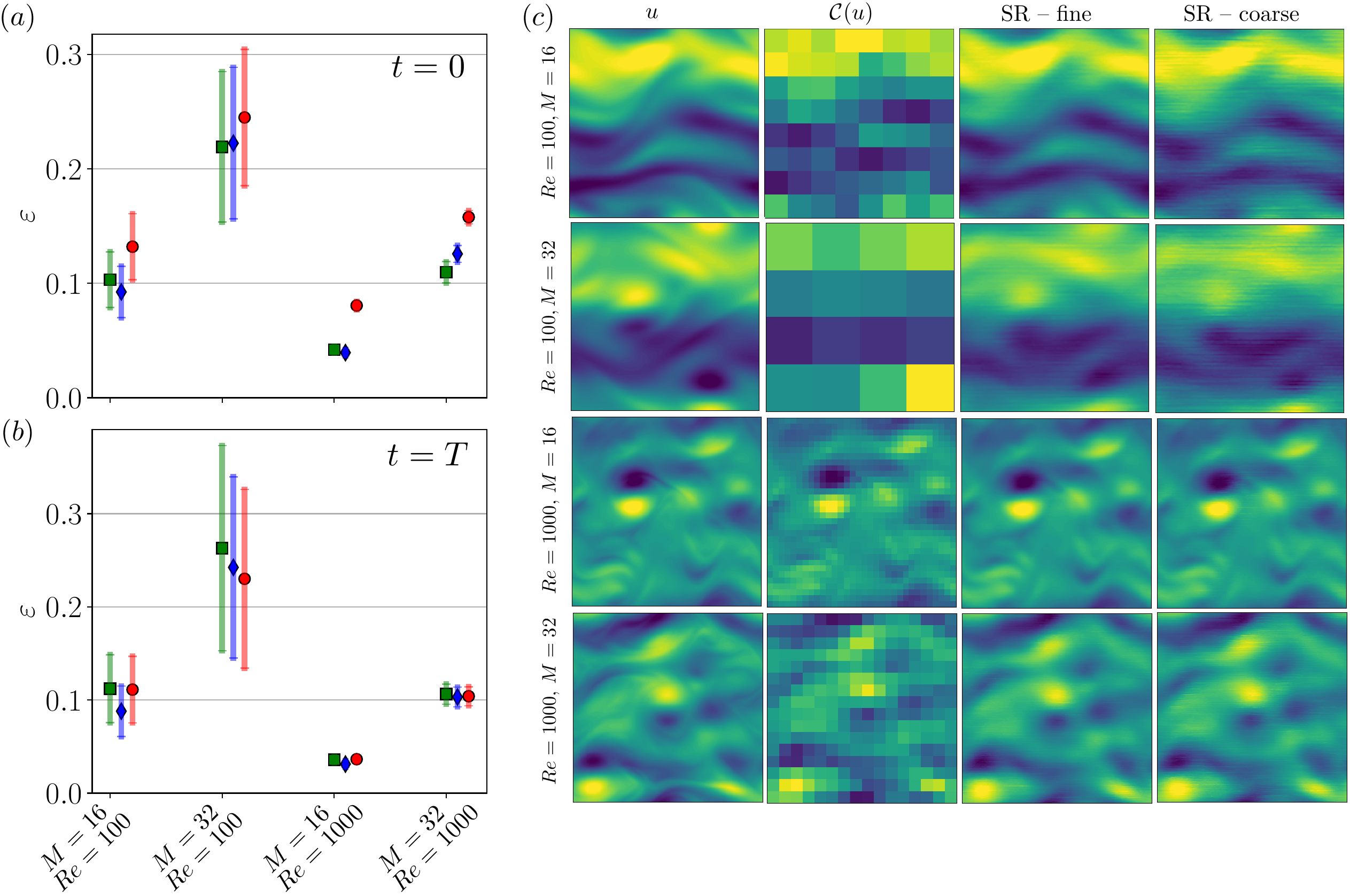}
    \caption{Summary of network performance at both $Re=100$ and $Re=1000$.
    (a) Average test-set errors (symbols; lines show $\pm$ a standard deviation). Models trained using `standard' super resolution loss (\ref{eqn:standard_sr_loss} are shown in green, time-advancement loss on the high-resolution grid (\ref{eqn:loss_nn_t}) in blue and the coarse-only time-dependent loss (\ref{eqn:loss_nn_da}) in red.
    (b) As (a) but errors are now computed after advancing ground truth and predictions forward in time by $T$.
    (c) Example model performance for a single snapshot at both $Re$ and coarsening factors. Streamwise velocity is shown with contours running between $\pm 3$. The output of models trained with losses (\ref{eqn:loss_nn_t}) and (\ref{eqn:loss_nn_da}) is shown.}
    \label{fig:network_summary_stats}
\end{figure}
The performance of the `dynamic' loss functions (\ref{eqn:loss_nn_t} and \ref{eqn:loss_nn_da}) is now assessed.
The test-average (and standard deviation) of the snapshot reconstruction error,
\begin{equation}
    \varepsilon_j(t) = \frac{\| \boldsymbol{\varphi}_t(\mathbf u_j) - \boldsymbol \varphi_t \circ \mathcal N_{\boldsymbol{\Theta}} \circ \mathcal C(\mathbf u_j)\|}{\| \boldsymbol{\varphi}_t(\mathbf u_j)\|},
\end{equation}
is reported both at $t=0$ and $t=T$ for all models in figure \ref{fig:network_summary_stats}.
The unroll time $T=1.5$ for all results at $Re=100$, while at $Re=1000$ two values are used; $T=0.5$ when $M=16$ and $T=1$ when $M=32$. 
Other models were trained with various similar $T$-values (not shown) and those presented here were found to be the best performing. 

Figure \ref{fig:network_summary_stats} includes results obtained with standard super-resolution (loss function \ref{eqn:standard_sr_loss}) along with both dynamic loss functions (\ref{eqn:loss_nn_t} and \ref{eqn:loss_nn_da}). 
When high-resolution data is available, the average errors in figure \ref{fig:network_summary_stats} indicate that including time evolution in the loss (\ref{eqn:loss_nn_t}) leads a marginal increase in reconstruction accuracy relative to the `standard' super-resolution approach in most cases.
This result holds for both $Re$ values and levels of coarse-graining. 
The error after the model outputs have been unrolled by $T$ is comparable to that at initial time for all coarsenings and $Re$ values. 

The $Re=100$ results are noticeably worse than those at $Re=1000$. 
This is perhaps unsurprising given that coarse-graining at the lower $Re$ value -- particularly at $M=32$ -- means that observations are available on a scale which is larger than typical vortical structures and much larger than the `critical' length $l_C$; see figures \ref{fig:vort_examples} and \ref{fig:network_summary_stats}(c). 
In contrast, the $M=32$ coarsening at $Re=1000$ is not as significant relative to $l_C$, though the spacing is still larger than this value. 

Most interestingly, the performance of the models trained \emph{without} high resolution data (loss function \ref{eqn:loss_nn_da}, red symbols in figure \ref{fig:network_summary_stats}), is also strong. 
The reconstruction errors at initial time $t=0$ are comparable to those obtained with high resolution models, e.g. at $Re=1000$ the $M=32$ error is $\sim 1.5 \times$ that from the high resolution models. 
A visual inspection of the reconstruction shows faithful reproduction of the larger scale features in the flow.
Examples of this behaviour are included in figure \ref{fig:network_summary_stats}(c),
where it is challenging to distinguish the differences by eye between the coarse and fine models at $Re=1000$.
The reconstruction errors decrease as the predictions from the coarse models are marched forward in time and at $t=T$ are near-identical to -- and sometimes lower than -- those obtained via the high-resolution approaches.

The drop in $\varepsilon(T)$ relative to $\varepsilon(0)$ for the coarse-grained-only approach at $Re=1000$ mimics observations of variational data-assimilation in turbulent flows \citep{Li2019,Wang2021}.
Data-assimilation is also considered here as a point of comparison for the performance of models trained only on coarse observations. 
Given coarse-grained observations of a ground truth trajectory $\{\mathcal C \circ \boldsymbol{\varphi}_{t_k}(\mathbf u_0)\}_{k=0}^{N_T-1}$, an initial condition $\mathbf v_0$ is sought to minimise the following loss function:
\begin{equation}
    \mathscr L_{DA}(\mathbf v_0) := \frac{1}{N_T}\sum_{k=1}^{N_T} \| \mathcal C \circ {\boldsymbol\varphi_{t_k}(\mathbf u_0)} - 
    \mathcal C \circ \boldsymbol\varphi_{t_k}(\mathbf v_0)
    \|^2.
    \label{eqn:loss_da}
\end{equation}
Equation (\ref{eqn:loss_da}) is minimised here using an Adam optimiser \citep{Kingma2015} in an approach similar to the training of neural networks outlined above. 
Gradients are evaluated automatically using the differentiablilty of the underlying solver rather than by construction of an adjoint system. 
Since data assimilation is performed on velocity (rather than the vorticity), gradients are projected onto a divergence-free solution prior to updating the initial condition. 
An initial learning rate of $\eta=0.2$ is used throughout and the number of iterations is fixed at $N_{it}=50$. 
The initial condition for the optimiser is the coarse-grained field $\mathcal C(\mathbf u_0)$ interpolated to the high-resolution grid with bicubic interpolation.


\begin{figure}
    \centering
    \includegraphics[width=0.9\linewidth]{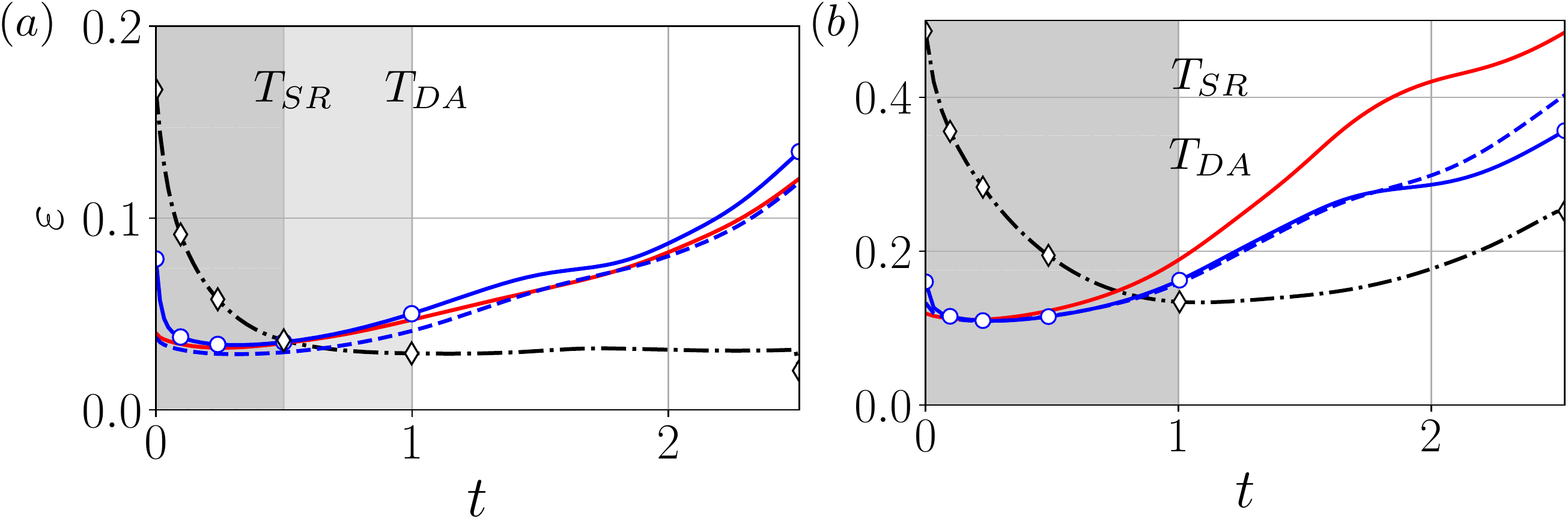}
    \caption{Comparison of model predictions under time advancement at $Re=1000$ to variational data assimilation for example evolutions at both $M=16$ (left) and $M=32$ (right). Black lines are the time-evolved predictions from data assimilation, red lines the performance of `standard' super-resolution, while blue lines show the performance of the time-dependent loss functions (\ref{eqn:loss_nn_t} and \ref{eqn:loss_nn_da}). Solid blue is the coarse-only version, and the symbols identify times where the fields are visualised in figures \ref{fig:Re1000_16assim} and \ref{fig:Re1000_32assim}.
    Grey regions highlight the assimilation window, where $T_{DA}=1$, and the unroll times to train the networks, $T=0.5$ at $M=16$ and $T=1$ at $M=32$.
    }
    \label{fig:Re1000_assim_1D}
\end{figure}
Example evolutions of the reconstruction error at $Re = 1000$ and both $M\in \{16, 32\}$ are reported in figure \ref{fig:Re1000_assim_1D} for networks trained with all loss functions (\ref{eqn:standard_sr_loss}--\ref{eqn:loss_nn_da}), as well as for the data-assimilation described above. 
As expected based on the statistics in figure \ref{fig:network_summary_stats}, the error between the coarse-only neural network output and the ground truth drops rapidly when time marching, becoming nearly indistinguishable from the networks trained on high-resolution data. 

Notably, the neural networks outperform data-assimilation as state-estimation for the initial condition.
This is most significant at $M=32$ where the reconstruction error from assimilation is $>40\%$ \citep[the coarse-graining at $M=32$ means observations are spaced farther apart than the critical length $l_C$ and assimilation is expected to struggle, see][]{Lalescu2013, Li2019, Wang2021}.
At first glance, this is somewhat surprising since the assimilation is performed over the time windows highlighted in grey in figure \ref{fig:Re1000_assim_1D}, i.e. for this specific trajectory, while the neural network must make an estimation based only on a single coarse-grained snapshot. 
Presumably, the improved performance is a result of the model seeing a large number of coarse-grained evolutions during training and hence building a plausible representation of the solution manifold, while each data-assimilation computation is independent.
The assimilated field does lead to better reconstruction at later times and at $M=16$ remains close to ground truth beyond the assimilation window, though on average the super-resolution approach is more accurate for $t < T_{DA}$. 
The super-resolved field from the coarse-only network has a similar time evolution to the standard super-resolution approaches. 

\begin{figure}
    \centering
    \includegraphics[width=0.9\linewidth]{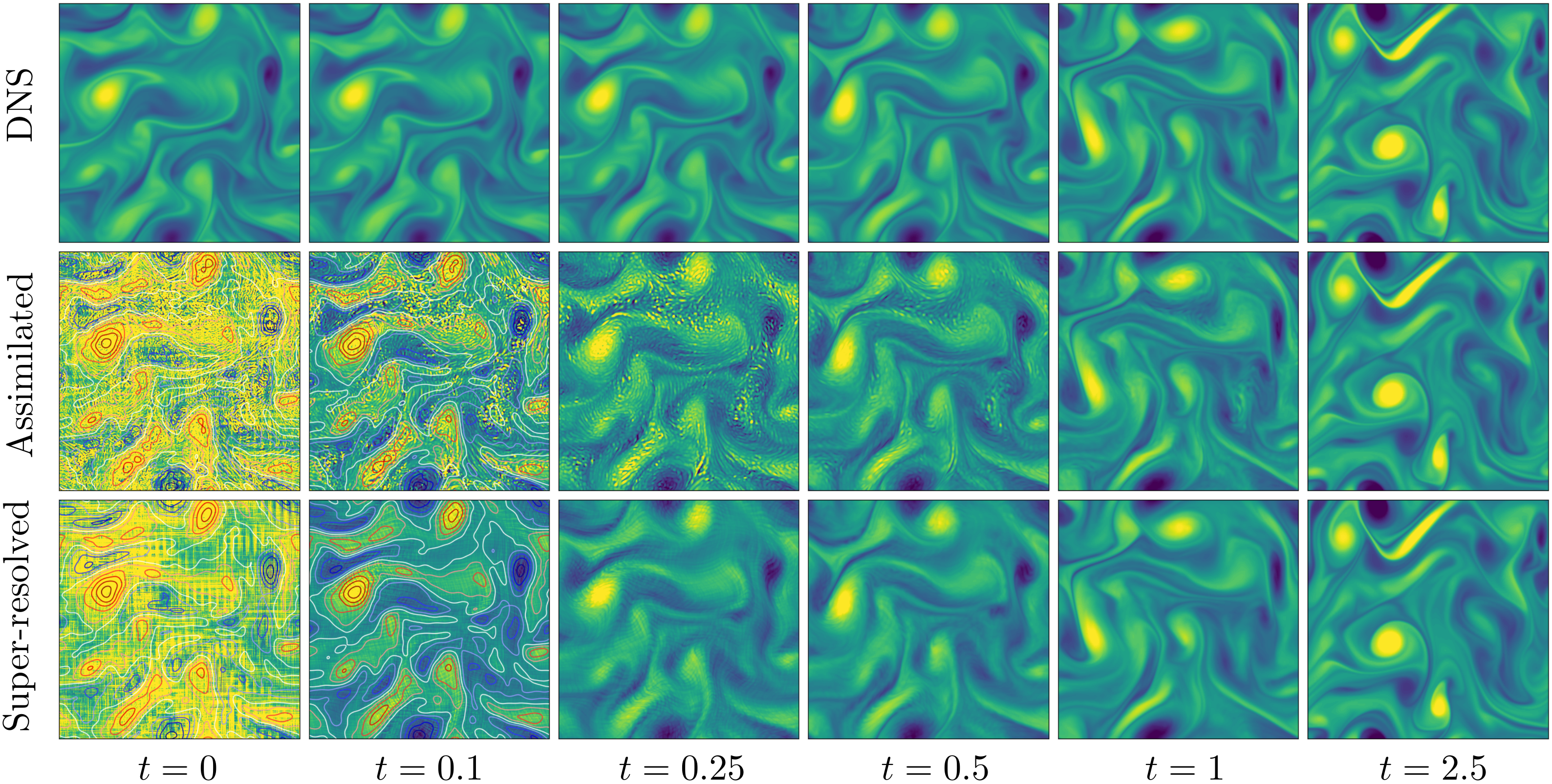}
    \caption{Evolution of the out-of-plane vorticity at $Re=1000$, above the evolution of the reconstructed field from data-assimilation and using the coarse-only super-resolution model (coarse graining here is $M=16$). Note the snapshots correspond to the trajectory reported in figure \ref{fig:Re1000_assim_1D} and are extracted at the times indicated by the symbols in that figure. Contours run between $\pm 15$. For $t\in \{0, 0.1\}$, a low-pass-filtered vorticity has been overlayed in red/blue lines to show the reproduction of the larger scale features which would otherwise be masked with small-scale noise. The cutoff wavenumber for the filter matches the coarse-graining and the contours are spaced by $\Delta \omega = 3$.}
    \label{fig:Re1000_16assim}
\end{figure}

A comparison of the time evolution of vorticity for the $M=16$ coarse-only model and the output of the data-assimilation optimisation is reported in figure \ref{fig:Re1000_16assim}.
In both approaches the initial voriticity is contaminated with high-wavenumber noise, though the low-wavenumber reconstruction is relatively robust (see additional line contours at early times in figure \ref{fig:Re1000_16assim}). 
For the neural network output, these small-scale errors quickly diffuse away and the vorticity closely matches the reference calculation by $t\sim 0.25$ (half of the unroll training time). 
In contrast, the assimilated field retains high-wavenumber error for much longer, although the large-scale structures are maintained on the whole.

\begin{figure}
    \centering
    \includegraphics[width=0.9\linewidth]{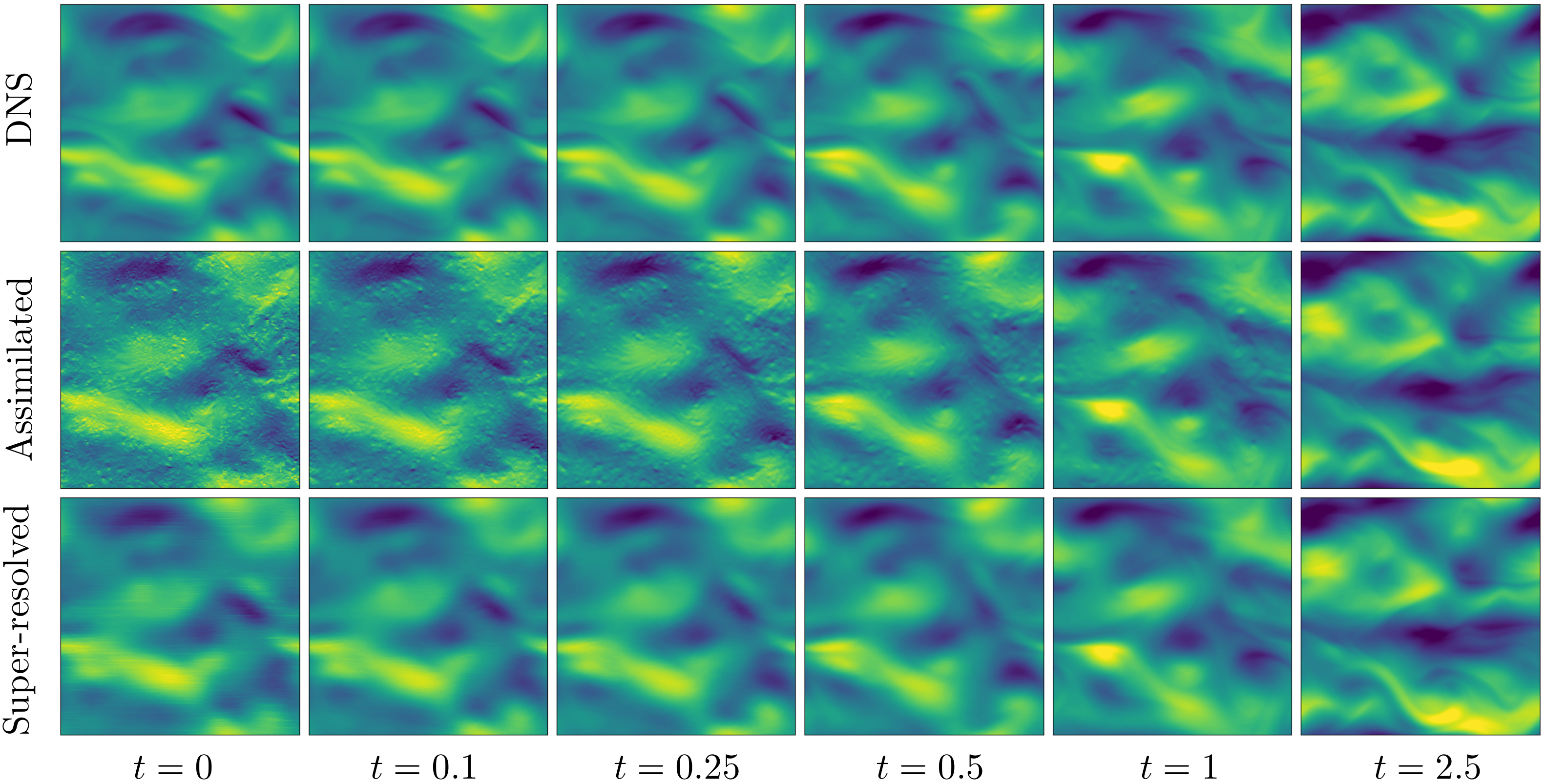}
    \caption{Evolution of streamwise velocity at $Re=1000$, above the evolution of the reconstructed field from data-assimilation and using the coarse-only super-resolution model (coarse graining here is $M=32$). Note the snapshots correspond to the trajectory reported in figure \ref{fig:Re1000_assim_1D} and are extracted at the times indicated by the symbols in that figure. Contour levels run between $\pm 3$.}
    \label{fig:Re1000_32assim}
\end{figure}

\begin{figure}
    \centering
    \includegraphics[width=0.8\linewidth]{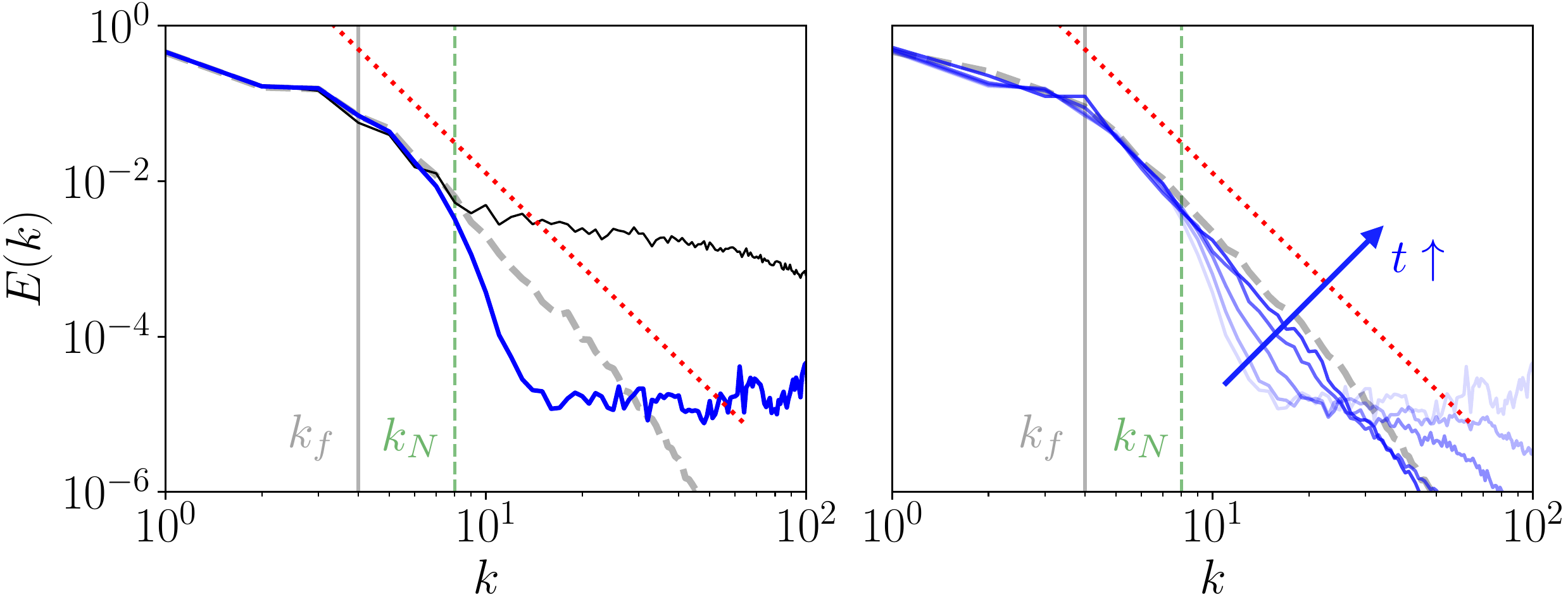}
    \caption{Energy spectra for coarse-grained network and assimilated fields.
    (Left) Energy in the initial condition for evolution in figure \ref{fig:Re1000_32assim} (grey), energy spectrum for the assimilated field (black) and super-resolved field (from coarse observations only, blue).
    (Right) Energy spectra in the time-advanced super-resolved field (blue) at times indicated in figure \ref{fig:Re1000_32assim} up to and including $t = 1$. Time-averaged energy spectrum of the true evolution is shown with the grey dashed line. 
    Vertical lines indicate the forcing wavenumber $k_f=4$ and the Nyquist cutoff wavenumber associated with the filter. Red lines show the scaling $E(k) \propto k^{-4}$.}
    \label{fig:Re1000_32spectra}
\end{figure}
These trends are exaggerated at the coarser value of $M=32$, as shown in $u$-evolution reported in figure \ref{fig:Re1000_32assim}. 
Again, the coarse-only neural network faithfully reproduces the larger-scale structures in the flow, with agreement in the smaller scales emerging under time marching. 
In contrast, the assimilated field remains contaminated by higher-wavenumber noise throughout the early evolution.
This difference in spectral content between the coarse network and assimilation is examined in figure \ref{fig:Re1000_32spectra}:
the super-resolved field better captures the low-wavenumber amplitudes before the Nyquist cutoff associated with the coarse-graining, while the assimilated field matches the scaling of the direct cascade (here $\sim k^{-4}$) over a slightly wider range of scales than the network prediction, but with significant energy in the small scales. 
In contrast, the super-resolved field has much lower-amplitude small scale noise; the spectrum rapidly adjusts to match ground truth under time evolution.

\section{Conclusions}

In this paper an approach to super-resolution has been presented which is similar in spirit to variational data assimilation. 
While the neural networks do a similar task to `standard' super-resolution, i.e. they take in a single snapshot of coarse-grained observations and predict a corresponding high-resolution field, the models are trained by requiring that the output of the network can be marched in time while remaining close to the target trajectory.
One consequence of this training procedure is that robust models can be trained without seeing high-resolution data -- they are trained to match the evolution of the coarse observations only.
The resulting models perform comparably to networks trained with high-resolution data, and can outperform data assimilation at state estimation. 


While these results represent a promising approach to state-estimation in situations where only coarse observations are available, they still rely on training models on relatively large amounts of data. 
Exploring the low-data-limit is an important next step.
There may also be scope to combine the super-resolution predictions with to better initialise 4DVar -- this has already shown promise with a learned inverse operator trained on high resolution data \citep{frerix2021}. 
Finally, the training procedure on coarse-only observations could be modified to try and improve the spectral content of the initial conditions. 
While such contamination is expected in data-assimilation, it may be relatively easy to address here by incorporating appropriate terms in the loss function.



\vspace{0.5cm}
\noindent
\textbf{Declaration of Interests.} The author reports no conflict of interest.

\vspace{0.5cm}
\noindent
\textbf{Acknowledgements.} Support from UKRI Frontier Guarantee Grant EP/Y004094/1 is gratefully acknowledged, as is support from the Edinburgh International Data Facility (EIDF) and the Data-Driven Innovation Programme at the University of Edinburgh.


\bibliographystyle{jfm}

\begin{thebibliography}{22}
\expandafter\ifx\csname natexlab\endcsname\relax\def\natexlab#1{#1}\fi
\def\au#1{#1} \def\ed#1{#1} \def\yr#1{#1}\def\at#1{#1}\def\jt#1{\textit{#1}}
  \def\bt#1{#1}\def\bvol#1{\textbf{#1}} \def\vol#1{#1} \def\pg#1{#1}
  \def\publ#1{#1}\def\arxiv#1{#1}\def\org#1{#1}\def\st#1{\textit{#1}}

\bibitem[Brunton {\em et~al.\/}(2020)Brunton, Noack \&
  Koumoutsakos]{Brunton2020}
{\sc \au{Brunton, Steven~L.}, \au{Noack, Bernd~R.} \& \au{Koumoutsakos,
  Petros}} \yr{2020}  \at{Machine learning for fluid mechanics}.  \jt{Annual
  Review of Fluid Mechanics}  \bvol{52}~(1),  \pg{477--508}.

\bibitem[Chandler \& Kerswell(2013)]{Chandler2013}
{\sc \au{Chandler, G.~J.} \& \au{Kerswell, R.~R.}} \yr{2013}  \at{Invariant
  recurrent solutions embedded in a turbulent two-dimensional {K}olmogorov
  flow}.  \jt{Journal of Fluid Mechanics}  \bvol{722},  \pg{554–595}.

\bibitem[Dong {\em et~al.\/}(2016)Dong, Loy, He \& Tang]{Dong2016}
{\sc \au{Dong, Chao}, \au{Loy, Chen~Change}, \au{He, Kaiming} \& \au{Tang,
  Xiaoou}} \yr{2016}  \at{Image super-resolution using deep convolutional
  networks}.  \jt{IEEE Transactions on Pattern Analysis and Machine
  Intelligence}  \bvol{38}~(2),  \pg{295–307}.

\bibitem[Dresdner {\em et~al.\/}(2022)Dresdner, Kochkov, Norgaard,
  Zepeda-Núñez, Smith, Brenner \& Hoyer]{Dresdner2022}
{\sc \au{Dresdner, Gideon}, \au{Kochkov, Dmitrii}, \au{Norgaard, Peter},
  \au{Zepeda-Núñez, Leonardo}, \au{Smith, Jamie~A.}, \au{Brenner, Michael~P.}
  \& \au{Hoyer, Stephan}} \yr{2022}  \at{Learning to correct spectral methods
  for simulating turbulent flows} .

\bibitem[Frerix {\em et~al.\/}(2021)Frerix, Kochkov, Smith, Cremers, Brenner \&
  Hoyer]{frerix2021}
{\sc \au{Frerix, Thomas}, \au{Kochkov, Dmitrii}, \au{Smith, Jamie},
  \au{Cremers, Daniel}, \au{Brenner, Michael} \& \au{Hoyer, Stephan}} \yr{2021}
  Variational data assimilation with a learned inverse observation operator.
  \bt{In {\em Proceedings of the 38th International Conference on Machine
  Learning\/} (ed. \ed{Marina Meila \& Tong Zhang})},  \st{Proceedings of
  Machine Learning Research},  \vol{vol. 139},  \pg{pp. 3449--3458}.
  \publ{PMLR}.

\bibitem[Fukami {\em et~al.\/}(2019)Fukami, Fukagata \& Taira]{Fukami2019}
{\sc \au{Fukami, Kai}, \au{Fukagata, Koji} \& \au{Taira, Kunihiko}} \yr{2019}
  \at{Super-resolution reconstruction of turbulent flows with machine
  learning}.  \jt{Journal of Fluid Mechanics}  \bvol{870},  \pg{106–120}.

\bibitem[Fukami {\em et~al.\/}(2020)Fukami, Fukagata \& Taira]{Fukami2020}
{\sc \au{Fukami, Kai}, \au{Fukagata, Koji} \& \au{Taira, Kunihiko}} \yr{2020}
  \at{Machine-learning-based spatio-temporal super resolution reconstruction of
  turbulent flows}.  \jt{Journal of Fluid Mechanics}  \bvol{909}.

\bibitem[Fukami {\em et~al.\/}(2023)Fukami, Fukagata \& Taira]{Fukami2023}
{\sc \au{Fukami, Kai}, \au{Fukagata, Koji} \& \au{Taira, Kunihiko}} \yr{2023}
  \at{Super-resolution analysis via machine learning: a survey for fluid
  flows}.  \jt{Theoretical and Computational Fluid Dynamics}  \bvol{37}~(4),
  \pg{421–444}.

\bibitem[Fukami \& Taira(2024)]{Fukami2024}
{\sc \au{Fukami, Kai} \& \au{Taira, Kunihiko}} \yr{2024} Single-snapshot
  machine learning for turbulence super resolution.

\bibitem[He {\em et~al.\/}(2016)He, Zhang, Ren \& Sun]{He2016}
{\sc \au{He, Kaiming}, \au{Zhang, Xiangyu}, \au{Ren, Shaoqing} \& \au{Sun,
  Jian}} \yr{2016} Deep residual learning for image recognition.  \bt{In {\em
  2016 IEEE Conference on Computer Vision and Pattern Recognition (CVPR)\/}}.
  \publ{IEEE}.

\bibitem[Hendrycks \& Gimpel(2016)]{gelu_arxiv}
{\sc \au{Hendrycks, Dan} \& \au{Gimpel, Kevin}} \yr{2016} Gaussian error linear
  units (gelus).

\bibitem[Kelshaw {\em et~al.\/}(2022)Kelshaw, Rigas \& Magri]{Kelshaw2022}
{\sc \au{Kelshaw, Daniel}, \au{Rigas, Georgios} \& \au{Magri, Luca}} \yr{2022}
  Physics-informed cnns for super-resolution of sparse observations on
  dynamical systems.  \bt{In {\em NeurIPS 2022 Workshop on Machine Learning and
  the Physical Sciences\/}}.

\bibitem[Kingma \& Ba(2015)]{Kingma2015}
{\sc \au{Kingma, Diederik~P.} \& \au{Ba, Jimmy}} \yr{2015} Adam: {A} method for
  stochastic optimization.  \bt{In {\em 3rd International Conference on
  Learning Representations, {ICLR} 2015, San Diego, CA, USA, May 7-9, 2015,
  Conference Track Proceedings\/} (ed. \ed{Yoshua Bengio \& Yann LeCun})}.

\bibitem[Kochkov {\em et~al.\/}(2021)Kochkov, Smith, Alieva, Wang, Brenner \&
  Hoyer]{Kochkov2021}
{\sc \au{Kochkov, D.}, \au{Smith, J.~A.}, \au{Alieva, A.}, \au{Wang, Q.},
  \au{Brenner, M.~P.} \& \au{Hoyer, S.}} \yr{2021}  \at{Machine
  learning-accelerated computational fluid dynamics}.  \jt{Proceedings of the
  National Academy of Sciences}  \bvol{118},  \pg{e2101784118}.

\bibitem[Lalescu {\em et~al.\/}(2013)Lalescu, Meneveau \& Eyink]{Lalescu2013}
{\sc \au{Lalescu, Cristian~C.}, \au{Meneveau, Charles} \& \au{Eyink,
  Gregory~L.}} \yr{2013}  \at{Synchronization of chaos in fully developed
  turbulence}.  \jt{Physical Review Letters}  \bvol{110}~(8).

\bibitem[Li {\em et~al.\/}(2019)Li, Zhang, Dong \& Abdullah]{Li2019}
{\sc \au{Li, Yi}, \au{Zhang, Jianlei}, \au{Dong, Gang} \& \au{Abdullah,
  Naseer~S.}} \yr{2019}  \at{Small-scale reconstruction in three-dimensional
  kolmogorov flows using four-dimensional variational data assimilation}.
  \jt{Journal of Fluid Mechanics}  \bvol{885}.

\bibitem[List {\em et~al.\/}(2022)List, Chen \& Thuerey]{List2022}
{\sc \au{List, Bj\"{o}rn}, \au{Chen, Li-Wei} \& \au{Thuerey, Nils}} \yr{2022}
  \at{Learned turbulence modelling with differentiable fluid solvers:
  physics-based loss functions and optimisation horizons}.  \jt{Journal of
  Fluid Mechanics}  \bvol{949}.

\bibitem[Page {\em et~al.\/}(2021)Page, Brenner \& Kerswell]{Page2021}
{\sc \au{Page, J.}, \au{Brenner, M.~P.} \& \au{Kerswell, R.~R.}} \yr{2021}
  \at{Revealing the state space of turbulence using machine learning}.
  \jt{Physical Review Fluids}  \bvol{6},  \pg{034402}.

\bibitem[Page {\em et~al.\/}(2024)Page, Norgaard, Brenner \&
  Kerswell]{Page2024}
{\sc \au{Page, Jacob}, \au{Norgaard, Peter}, \au{Brenner, Michael~P.} \&
  \au{Kerswell, Rich~R.}} \yr{2024}  \at{Recurrent flow patterns as a basis for
  two-dimensional turbulence: Predicting statistics from structures}.
  \jt{Proceedings of the National Academy of Sciences}  \bvol{121}~(23).

\bibitem[Um {\em et~al.\/}(2020)Um, Brand, Fei, Holl \& Thuerey]{solverinloop}
{\sc \au{Um, Kiwon}, \au{Brand, Robert}, \au{Fei, Yun~(Raymond)}, \au{Holl,
  Philipp} \& \au{Thuerey, Nils}} \yr{2020} Solver-in-the-loop: Learning from
  differentiable physics to interact with iterative pde-solvers.  \bt{In {\em
  Advances in Neural Information Processing Systems\/} (ed. \ed{H.~Larochelle,
  M.~Ranzato, R.~Hadsell, M.F. Balcan \& H.~Lin})}, ,  \vol{vol.~33},  \pg{pp.
  6111--6122}.  \publ{Curran Associates, Inc.}

\bibitem[Wang \& Zaki(2021)]{Wang2021}
{\sc \au{Wang, Mengze} \& \au{Zaki, Tamer~A.}} \yr{2021}  \at{State estimation
  in turbulent channel flow from limited observations}.  \jt{Journal of Fluid
  Mechanics}  \bvol{917}.

\bibitem[Wang \& Zaki(2022)]{Wang2022}
{\sc \au{Wang, Mengze} \& \au{Zaki, Tamer~A.}} \yr{2022}  \at{Synchronization
  of turbulence in channel flow}.  \jt{Journal of Fluid Mechanics}  \bvol{943}.

\end{thebibliography}

\end{document}